\documentclass{article}
\usepackage[latin9]{inputenc}
\usepackage{amsmath}
\usepackage{graphicx}

\makeatletter

\providecommand{\tabularnewline}{\\}

\usepackage{spconf}

\usepackage[pdftex,pdftitle={Low-Complexity, Real-Time Joint Neural Echo Control and Speech Enhancement Based On PercepNet},pdfauthor={Jean-Marc Valin, Srikanth Tenneti, Karim Helwani, Umut Isik, Arvindh Krishnaswamy}]{hyperref}
\usepackage{balance}

\makeatother

\begin{document}
\name{Jean-Marc Valin, Srikanth Tenneti, Karim Helwani, Umut Isik, Arvindh Krishnaswamy} \address{Amazon Web Services\\ Palo Alto, CA, USA\\ \small{\texttt{\{jmvalin, stenneti, helwk, umutisik, arvindhk\}@amazon.com}}}
\ninept

\title{Low-Complexity, Real-Time Joint Neural Echo Control and Speech Enhancement
Based On PercepNet}

\maketitle
\maketitle 
\begin{abstract}
Speech enhancement algorithms based on deep learning have greatly
surpassed their traditional counterparts and are now being considered
for the task of removing acoustic echo from hands-free communication
systems. This is a challenging problem due to both real-world constraints
like loudspeaker non-linearities, and to limited compute capabilities
in some communication systems. In this work, we propose a system combining
a traditional acoustic echo canceller, and a low-complexity joint
residual echo and noise suppressor based on a hybrid signal processing/deep
neural network (DSP/DNN) approach. We show that the proposed system
outperforms both traditional and other neural approaches, while requiring
only 5.5\% CPU for real-time operation. We further show that the system
can scale to even lower complexity levels.
\end{abstract}
\begin{keywords}acoustic echo cancellation, neural residual echo
suppression, speech enhancement\end{keywords}

\section{Introduction}

\label{sec:intro}

In full-duplex communication applications, echo produced by the acoustic
feedback from the loudspeaker to the microphone can severely degrade
quality. Traditional acoustic echo cancellation (AEC) aims at cancelling
the acoustic echoes from the microphone signal by filtering the far-end
(loudspeaker) signal with the estimated echo path modeled by an adaptive
FIR filter, and subtracting the resulting signal from the microphone
signal~\cite{haykin96,buchner2003}. If the estimated echo path is
equal to the true echo path, echo is removed from the microphone signal.
In real-world applications, residual echo remains at the output of
AEC due to issues such as non-linearities in the acoustic drivers,
rapidly-varying acoustic environments, and microphone noise. Hence,
residual echo suppressors are typically employed after the system
identification-based AEC in order to meet the requirements for high
echo attenuation~\cite{faller06,Favrot12,Helwani13}.

In addition, background noise also degrades the speech quality, while
limiting the ability of the AEC to adapt fast enough to track acoustic
path changes, further worsening the overall communication quality.
Traditional speech enhancement methods~\cite{boll1979suppression,ephraim1985speech}
-- sometimes combined with acoustic echo suppression~\cite{gustafsson2002psychoacoustic}
-- can help reduce the effect of stationary noise, but have been mostly
unable to remove highly non-stationary noise. In recent years, deep-learning-based
speech enhancement systems have emerged as state-of-the-art solutions~\cite{xia2013speech,xu2014regression,weninger2015speech,isik2020poconet,reddy2020interspeech}.
Even more recently, deep-learning-based residual echo suppression
algorithms have also demonstrated state-of-the-art performance~\cite{ma2020acoustic,zhang2019deep}. 

In this paper, we present an integrated approach to noise suppression
and echo control (Section~\ref{sec:Overview}) which abides to the
idea of incorporating prior knowledge from physics and psychoacoustics
to design a low complexity but effective architecture. Since, the
acoustic path between a loudspeaker and a microphone is well approximated
as a linear FIR filter, we retain the traditional frequency-domain
acoustic echo canceller (AEC) described in Section~\ref{sec:adaptive-fiter}.
We combine the adaptive filter with a perceptually-motivated joint
noise and echo suppression algorithm (Section~\ref{sec:residual-echo-suppression}).
As in~\cite{valin2020perceptually}, we focus on restoring the spectral
envelope and the periodicity of the speech. Our model is trained (Section~\ref{sec:DNN-model})
to enhance the speech from the AEC using the far-end signal as side
information to help remove the far-end signal while denoising the
near-end speech. Results from our experiments and from the Acoustic
Echo Cancellation Challenge~\cite{Sridar2020} show that the proposed
algorithm outperforms both traditional and other neural approaches
to residual echo suppression, taking first place in the challenge
(Section~\ref{sec:Results}).

\section{Signal Model}

\label{sec:Overview}

\begin{figure}
\centering{\includegraphics[width=0.8\columnwidth]{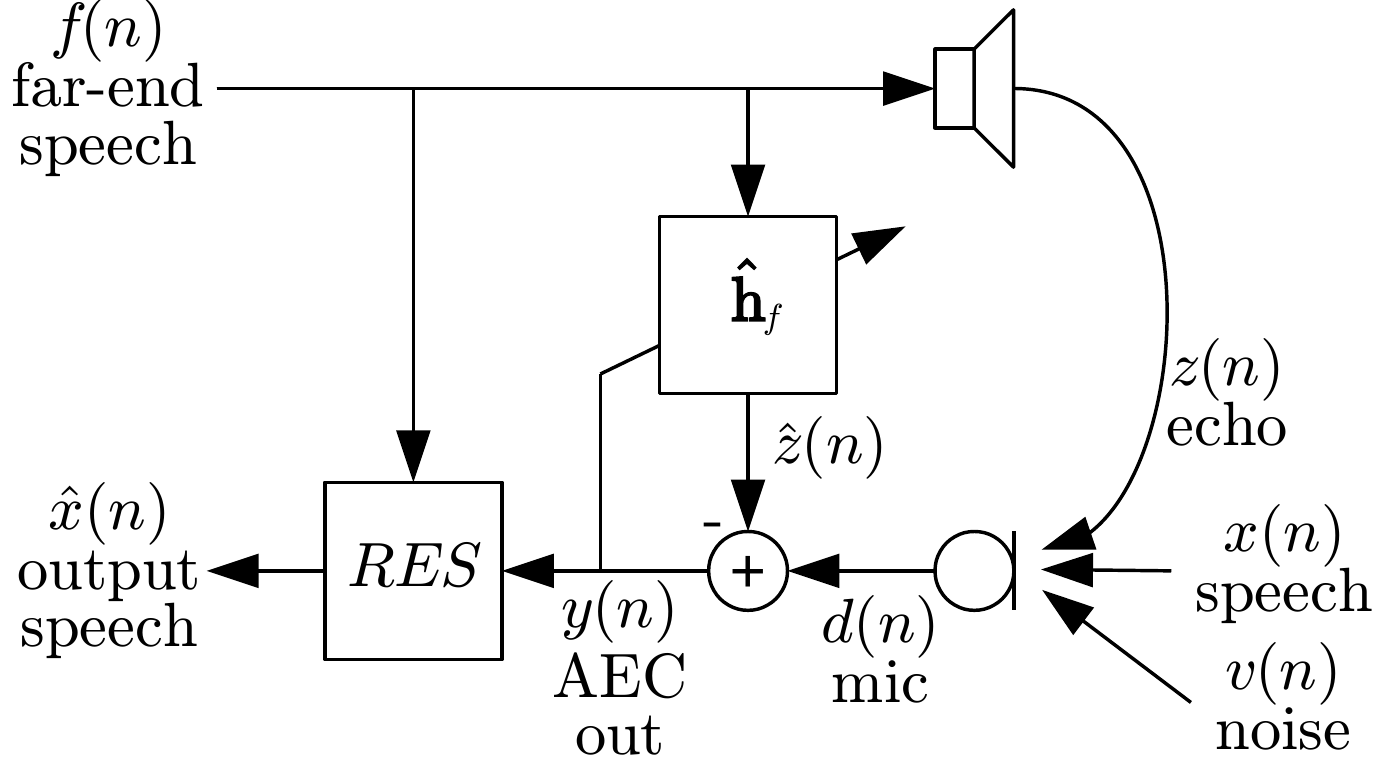}}

\caption{Overview of the joint echo control and noise suppression system. The
far-end signal $f\left(n\right)$ is played through the loudspeaker.
The microphone signal $d\left(n\right)$ captures the reverberated
near-end speech but also some noise $v\left(n\right)$, as well as
echo $z\left(n\right)$ from the loudspeaker. The echo is partially
cancelled by the adaptive filter $\hat{\mathbf{h}}_{f}$ to produce
$y\left(n\right)$. The RES then enhances $y\left(n\right)$ by suppressing
noise, reverberation, as well as the remaining echo, and produces
the enhanced output $\hat{x}\left(n\right)$.\label{fig:Signal-model}}
\end{figure}

The signal model we consider in this work is shown in Fig.~\ref{fig:Signal-model}.
Let $x\left(n\right)$ be a clean speech signal. The signal captured
by a hands-free microphone in a noisy room is given by
\begin{equation}
d\left(n\right)=x\left(n\right)\star\mathbf{h}_{x}+v\left(n\right)+z\left(n\right)\,,\label{eq:additive-noise}
\end{equation}
where $v\left(n\right)$ is the additive noise from the room, $z\left(n\right)$
is the echo caused by a far-end signal $f\left(n\right)$, $\mathbf{h}_{x}$
is the impulse response from the talker to the microphone, and $\star$~denotes
the convolution. When ignoring non-linear effects, the echo signal
can be expressed as $z\left(n\right)=f\left(n\right)\star\mathbf{h}_{f}$.
Echo cancellation based on adaptive filtering consists in estimating
$\mathbf{h}_{f}$ and subtracting the estimated echo $\hat{z}\left(n\right)$
from the microphone signal to produce the echo-cancelled signal $y\left(n\right)$.
Unfortunately, the echo cancellation process is generally imperfect
and echo remains in $y\left(n\right)$. For this reason, we include
a joint residual echo suppression (RES) and noise suppression (NS)
algorithm (RES block in Fig.~\ref{fig:Signal-model}) such that the
enhanced output $\hat{x}\left(n\right)$ is \emph{perceptually} as
close as possible to the ideal clean speech $x\left(n\right)$.

\section{Adaptive Filter}

\label{sec:adaptive-fiter}

The adaptive filter component in Fig.~\ref{fig:Signal-model} is
derived from the SpeexDSP\footnote{\url{https://gitlab.xiph.org/xiph/speexdsp/}}
implementation of the multidelay block frequency-domain (MDF) adaptive
filter~\cite{soo1990mdf} algorithm. Robustness to double-talk is
achieved through a combination of the learning rate control in~\cite{valin2007adjusting}
and a two-echo-path model as described in~\cite{ochiai1977echo}.
Moreover, a block variant of the PNLMS algorithm~\cite{duttweiler2000proportionate}
is used to speed up adaptation. As a compromise between complexity
and convergence, we use a variant of AUMDF~\cite{soo1990mdf} where
most blocks are alternatively constrained, but the highest-energy
block is constrained on each iteration.

There is sometimes an unknown delay between the signal $f\left(n\right)$
sent to the loudspeaker and the corresponding echo appearing at the
microphone. To estimate that delay $D$, we run a second AEC with
a 400\nobreakdash-ms filter and find the peak in the estimated filter.
The delay-estimating AEC operates on a down-sampled version of the
signals (8~kHz) to reduce complexity. We use the delayed far-end
signal $f\left(n-D\right)$ to perform the final echo cancellation
at 16~kHz. We use a frame size of 10~ms, which matches the frame
size used in the RES and avoids causing any extra delay. 

The length of the adaptive filter affects not only the complexity,
but also the convergence time and the steady-state accuracy of the
filter. We have found that a 150\nobreakdash-ms filter provides a
good compromise, ensuring that the echo loudness is sufficiently reduced
for the RES to correctly preserve double-talk. We do not make any
attempt at cancelling non-linear distortion in the echo. 

\begin{figure}
\centering{\includegraphics[width=1\columnwidth]{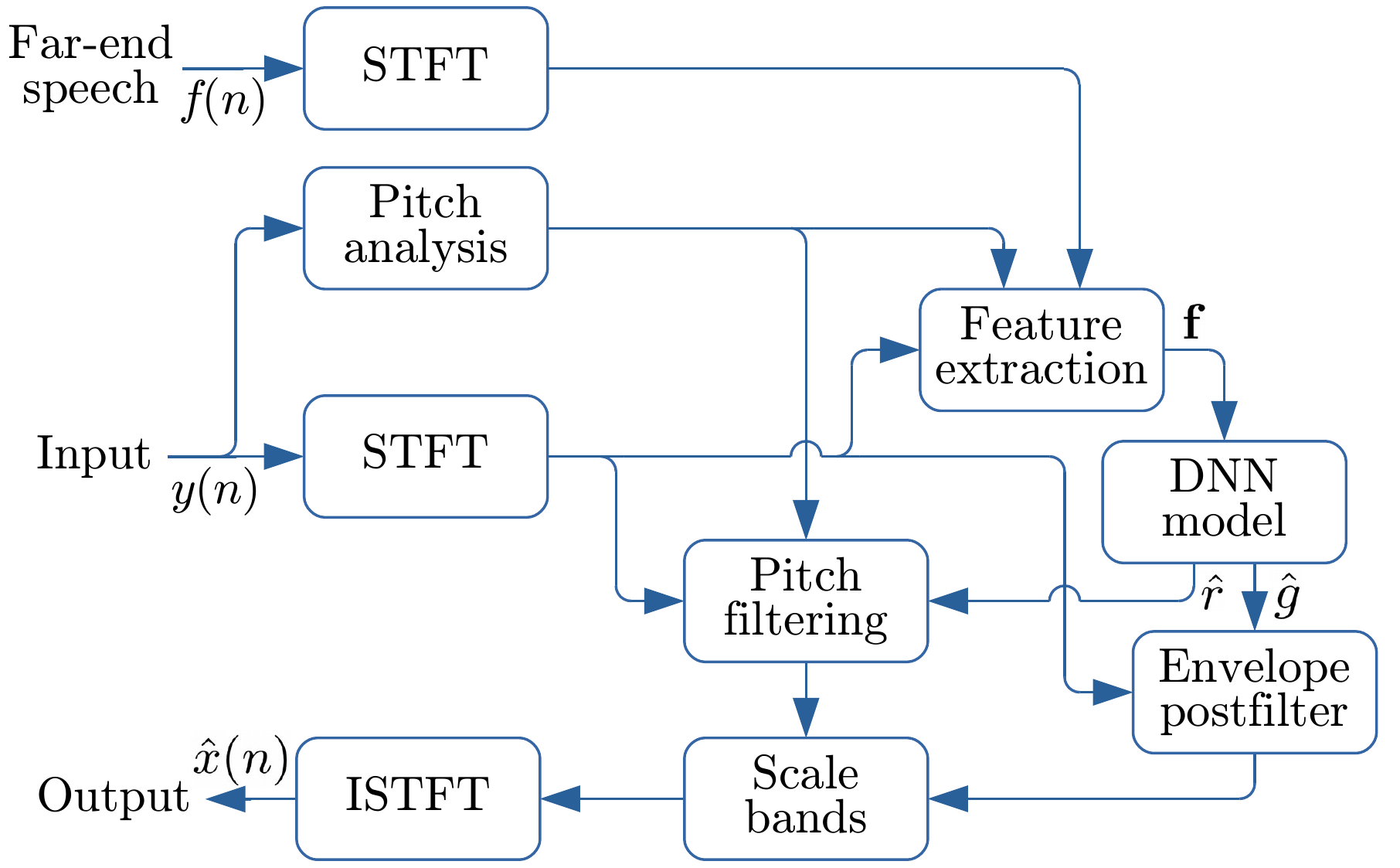}}\caption{Overview of the PercepNet joint noise and residual echo suppressor.\label{fig:PercepNet-overview}}
\end{figure}

\section{Residual Echo Suppression}

\label{sec:residual-echo-suppression}

The linear AEC output $y\left(n\right)$ contains the near-end speech
$x\left(n\right)$, the near-end noise $v\left(n\right)$, as well
as some residual echo $z\left(n\right)-\hat{z}\left(n\right)$. The
residual echo component includes
\begin{itemize}
\item misalignment (or divergence) of the estimated filter $\hat{\mathbf{h}}_{f}$
\item non-linear distortion caused by the loudspeaker
\item late reverberation beyond the impulse response of $\hat{\mathbf{h}}_{f}$
\end{itemize}
Unlike the problem of noise suppression, residual echo suppression
involves isolating a speech signal from another speech signal. Since
the echo can sometimes be indistinguishable from the near-end speech,
additional information is required for neural echo suppression to
work reliably. While there are multiple ways to provide information
about the echo, we have found that using the far-end signal $f\left(n\right)$
is both the simplest and the most effective way. Specifically, since
$f\left(n\right)$ does not depend on the AEC behaviour, convergence
problems with the echo canceller are less likely to affect the RES
performance. Similarly, we found that using the delayed signal $f\left(n-D\right)$
leads to slightly poorer results -- most likely due to the few cases
where delay estimation fails. 

We implement joint RES and NS using the PercepNet algorithm~\cite{valin2020perceptually},
which is based on two main ideas:
\begin{itemize}
\item scaling the energy of perceptually-spaced spectral bands to match
that of the near-end speech;
\item using a multi-tap comb filter at the pitch frequency to remove noise
between harmonics and match the periodicity of the near-end speech.
\end{itemize}
Let $Y_{b}\left(\ell\right)$ be the magnitude of the AEC output signal
$y\left(n\right)$ in band~$b$ for frame $\ell$ and $X_{b}\left(\ell\right)$
be similarly defined for the clean speech $x\left(n\right)$, the
ideal gain that should be applied to that band is:
\begin{equation}
g_{b}\left(\ell\right)=\frac{X_{b}\left(\ell\right)}{Y_{b}\left(\ell\right)}\,.\label{eq:magnitude-gain}
\end{equation}
Applying the gain $g_{b}\left(\ell\right)$ to the magnitude spectrum
in band $b$ results in an enhanced signal that has the same spectral
envelope as the clean speech. While this is generally sufficient for
unvoiced segments, voiced segment are likely have a higher \emph{roughness}
than the clean speech. This is due to noise between harmonics reducing
the perceived periodicity/voicing of the speech. The noise is particularly
perceptible due to the fact that tones have relatively little masking
effect on noise~\cite{gockel2003asymmetry}. In that situation, we
use a non-causal comb filter to remove the noise between the pitch
harmonics and make the signal more periodic. The comb filter is controlled
by strength/mixing parameters $r_{b}\left(\ell\right)$, where $r_{b}\left(\ell\right)=0$
causes no filtering to occur and $r_{b}\left(\ell\right)=1$ causes
the band to be replaced by the comb-filtered version, maximizing periodicity.
In cases where even $r_{b}\left(\ell\right)=1$ is insufficient to
make the noise inaudible, a further attenuation $g_{b}^{\left(\mathrm{att}\right)}\left(\ell\right)$
is applied (Section~3 of~\cite{valin2020perceptually}).

Fig.~\ref{fig:PercepNet-overview} shows an overview of the RES algorithm.
The short-time Fourier transform (STFT) spectrum is divided into 32~triangular
bands following the equivalent rectangular bandwidth (ERB) scale~\cite{moore2012introduction}.
The features computed from the input and far-end speech signals are
used by a deep neural network (DNN) to estimate the gains $\hat{g}_{b}\left(\ell\right)$
and filtering strengths $\hat{r}_{b}\left(\ell\right)$ to use. The
output gains $\hat{g}_{b}\left(\ell\right)$ are further modified
by an envelope postfilter (Section~5 of~\cite{valin2020perceptually})
that reduces the perceptual impact of the remaining noise in each
band.

\section{DNN model}

\label{sec:DNN-model}

\begin{figure}
\centering{\includegraphics[width=0.85\columnwidth]{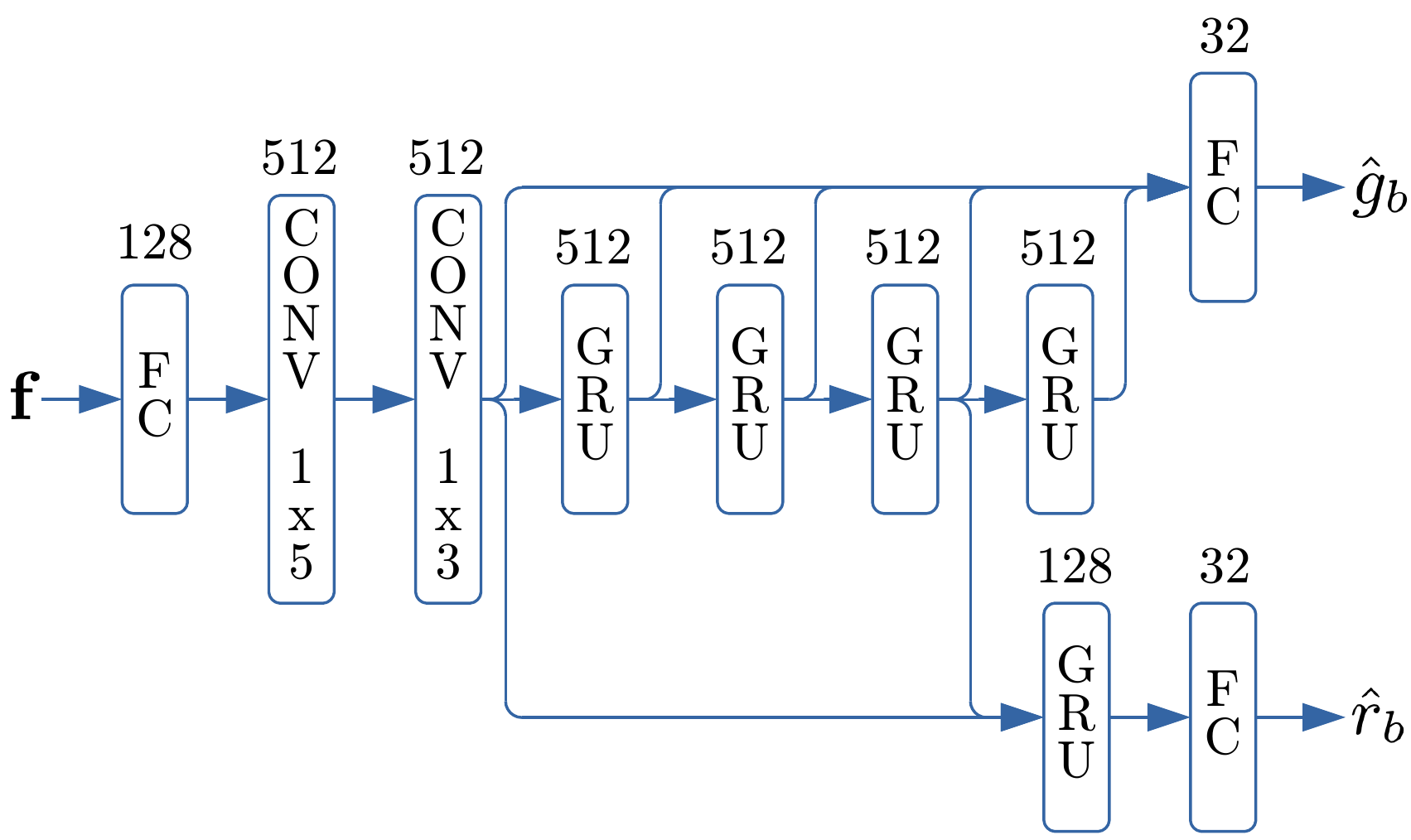}}\caption{Overview of the DNN architecture computing the 32~gains $\hat{g}_{b}$
and 32~strengths $\hat{r}_{b}$ from the 100\protect\nobreakdash-dimensional
input feature vector $\mathbf{f}$. The number of units on each layer
is indicated above the layer type.\label{fig:Overview-of-DNN}}
\end{figure}

The model uses two convolutional layers (a 1x5~layer followed by
a~1x3 layer), and five GRU~\cite{cho2014properties} layers, as
shown in Fig.~\ref{fig:Overview-of-DNN}. The convolutional layers
are aligned in time such as to use up to $M$ frames into the future.
In order to achieve the 40~ms algorithmic delay allowed by the challenge~\cite{Sridar2020},
including the 10\nobreakdash-ms frame size and the 10\nobreakdash-ms
overlap, we have $M=2$.

The input features used by the model are tied to the 32~bands we
use. For each band, we use three features:
\begin{enumerate}
\item the energy in the band with look-ahead $Y_{b}\left(\ell+M\right)$
\item the pitch coherence~\cite{valin2020perceptually} without look-ahead
$q_{y,b}\left(\ell\right)$ (the coherence estimation itself uses
the full look-ahead), and
\item the energy of the far-end band with look-ahead $F_{b}\left(\ell+M\right)$
\end{enumerate}
In addition to those 96~band-related features, we use four extra
scalar features (for a total of 100 input features):
\begin{itemize}
\item the pitch period~$T\left(\ell\right)$,
\item an estimate of the pitch correlation with look-ahead,
\item a non-stationarity estimate, and
\item the ratio of the $L_{1}$-norm to the $L_{2}$-norm of the excitation
computed from $y\left(n\right)$.
\end{itemize}
For each band $b$, we have 2~outputs: the gain $\hat{g}_{b}\left(\ell\right)$
approximates $g_{b}^{\left(\mathrm{att}\right)}\left(\ell\right)g_{b}\left(\ell\right)$
and the strength $\hat{r}_{b}\left(\ell\right)$ approximates $r_{b}\left(\ell\right)$. 

The 8M~weights in the model are forced to a $\pm\frac{1}{2}$ range
and quantized to 8-bit integers. This reduces the total memory requirement
(and cache bandwidth), while also reducing the computational complexity
of the inference when taking advantage of vectorization (more operations
for the same register width).

\subsection{Sparse model}

\label{sub:sparse-model}

In some situations, it is desirable to further reduce the complexity
of the model. While it is always possible to reduce the number of
units in each layer, it has recently been found that using sparse
weight matrices (i.e. sparse network connections) can lead to better
results~\cite{narang2017exploring,zhu2017prune}. Since modern CPUs
make heavy use of \emph{single instruction, multiple data} (SIMD)
hardware, it is important for the algorithm to allow vectorization.
For that reason, we use structured sparsity -- where whole sub-blocks
of matrices are chosen to be either zero or non-zero -- implemented
in a similar way to~\cite{kalchbrenner2018efficient,valin2019lpcnet}.
In this work, we use 16x4 sub-blocks. All fully-connected layers,
as well as the first convolutional layer are kept dense (no sparsity).
The second convolutional layer is 50\% dense, and the GRUs use different
levels of sparsity for the different gates. The matrices that compute
the new state have a density of 40\%, whereas the update gate matrices
are 20\% dense and the reset gate matrices have only 10\% density.
This reflects the unequal usefulness of the different gates on recurrent
units. 

The resulting sparse model has 2.1M non-zero weights, or 25\% of the
size of the full model. We also consider an even lower complexity
model with the same density but layers limited to 256~units, resulting
in 800k non-zero weights, or 10\% of the full model size. When training
sparse models, we use the sparsification schedule proposed in~\cite{zhu2017prune}.

\subsection{Training}

We train the model on synthetic mixtures of clean speech, noise and
echo that attempt to recreate real-world conditions, including reverberation.
We vary the signal-to-noise ratio (SNR) from -15~dB to 45 dB (with
some noise-free examples included), and the echo-to-near-end ratio
is between -15~dB and 35~dB. We use 120~hours of clean speech data
along with 80~hours of various noise types. Most of the data is sampled
at 48~kHz, but some of it -- including the far-end single-talk data
provided by the challenge organizers -- is sampled at 16~kHz. We
use both synthetic and real room impulse responses for the augmentation
process. 

In typical conditions, the effect of the room acoustics on the near-end
speech, the echo, and the noise is similar, but not identical. This
is due to the fact that while all three occur in the same room (same
$RT_{60}$), they can be in different locations and -- especially
-- at different distances. For that reason, we pick only one room
impulse response for each condition, but scale the early reflections
(first 20~ms) with a gain varying between 0.5 and 1.5 to simulate
the distance changing. Inspired by~\cite{zhao2018late}, the target
signal includes the early reflections as well as an attenuated echo
tail (with $RT_{60}=200\,\mathrm{ms}$) so that late reverberation
is attenuated to match the acoustics of a small room. 

We improve the generalization of the model by using various filtering
augmentation methods~\cite{valin2018rnnoise,valin2020perceptually}.
That includes applying a low-pass filter with a random cutoff frequency,
making it possible to use the same model on narrowband to fullband
audio. 

The loss function used for the gain attempts to match human perception
as closely as possible. For this reason we use the following loss
function for the gain estimations:
\begin{equation}
\mathcal{L}_{g}=\sum_{b}\mathcal{D}\left(g_{b},\hat{g}_{b}\right)+\lambda_{4}\sum_{b}\left[\mathcal{D}\left(g_{b},\hat{g}_{b}\right)\right]^{2}\ ,\label{eq:gain-loss}
\end{equation}
with the distortion function
\begin{equation}
\mathcal{D}\left(g_{b},\hat{g}_{b}\right)=\frac{\left(g_{b}^{2\gamma}-\hat{g}_{b}^{2\gamma}\right)^{2}}{\max\left(g_{b}^{2\gamma},\ \hat{g}_{b}^{2\gamma}\right)+\epsilon}\ ,\label{eq:gain-loss-dist}
\end{equation}
where $\gamma=0.3$ is the generally agreed-upon exponent to convert
acoustic power to the \emph{sone} scale for perceived loudness~\cite{moore2012introduction}.
The purpose of the denominator in \eqref{eq:gain-loss-dist} is to
over-emphasize the loss when completely attenuating speech or when
letting through small amounts of noise/echo during silence. We use
$\lambda_{4}=10$ for the second term of \eqref{eq:gain-loss}, an
$L_{4}$ term that over-emphasizes large errors in general. We use
the same loss function as~\cite{valin2020perceptually} for $\hat{r}_{b}$.

\section{Experiments and Results}

\label{sec:Results}

The complexity of the proposed RES with the largest (non-sparse) model
is dominated by the 800M~multiply-accumulate operations per second
required to compute the contribution of all 8M~weights on 100~frames
per second. The RES thus requires 4.6\% of an x86 mobile CPU core
(Intel i7-8565U) to operate in real-time. When combined with the AEC,
the total complexity of the proposed 16~kHz echo control solution
as submitted to the AEC challenge~\cite{Sridar2020} is 5.5\%~CPU
(0.55~ms per 10\nobreakdash-ms frame). Since the RES is already
designed to operate at 48~kHz, the total cost of fullband echo control
only increases to 6.6\%, with the difference due to the increased
AEC sampling rate.

\begin{table}
\caption{AEC Challenge official results: P.808 MOS of near-end single-talk,
P.831 Echo DMOS for far-end single-talk, P.831 Echo DMOS for double-talk,
P.831 other degradations DMOS of double-talk. The baseline model is
provided by the challenge organizers. As a comparison, we also include
the mean of the four algorithms statistically tied in second place.\label{tab:Challenge-official-P.808-MOS}}

\vspace{1em}

\centering{%
\begin{tabular}{lccccc}
\hline 
Algorithm & ST & ST & DT & DT & \textbf{Mean}\tabularnewline
 & NE & FE & Echo & Other & \tabularnewline
\hline 
Baseline & 3.79 & 3.84 & 3.84 & 3.28 & 3.68\tabularnewline
2$^{\mathrm{nd}}$ place & 3.80 & 4.18 & 4.25 & 3.74 & 3.99\tabularnewline
\textbf{PercepNet} & \textbf{3.85} & \textbf{4.19} & \textbf{4.34} & \textbf{4.07} & \textbf{4.11}\tabularnewline
\hline 
\end{tabular}}
\end{table}

The AEC challenge organizers evaluated \emph{blind} test samples processed
with the above AEC, followed by the PercepNet-based RES. The mean
opinion score (MOS)~\cite{P.800,P.831} results were obtained using
the crowdsourcing methodology described in P.808~\cite{P.808}. The
test set includes 1000~real recordings. Each utterance was rated
by 10~listeners, leading to a 95\% confidence interval of 0.01~MOS
for all algorithms. The proposed algorithm significantly out-performs
the ResRNN baseline, as shown in Table~\ref{tab:Challenge-official-P.808-MOS},
and ranked in first place among the 17~submissions to the challenge.
An interesting observation is that although the proposed algorithm
performs well over all the metrics, the improvement over the other
submitted algorithms is particularly noticeable for the ``DT Other''
metric, which measures the degradation caused to the near-end speech
during double-talk conditions. 

\begin{figure}
\centering{\includegraphics[width=1\columnwidth]{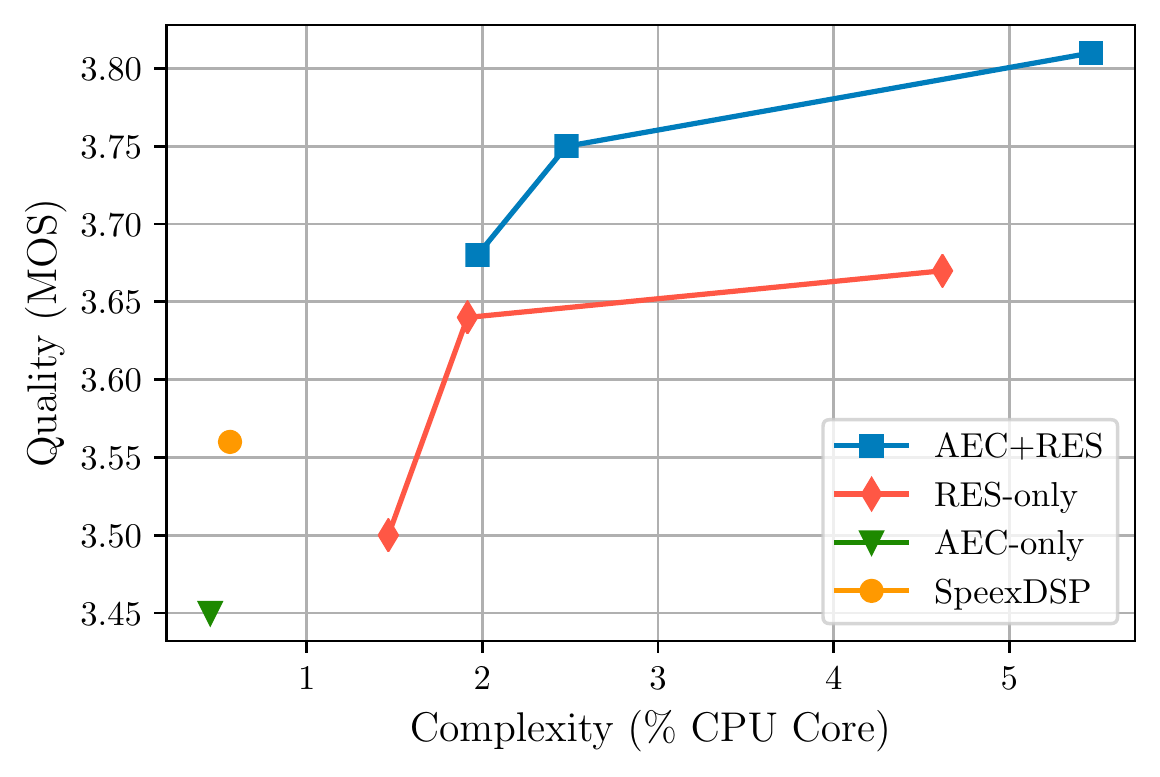}}

\caption{P.808 MOS results as a function of complexity. The 95\% confidence
interval is 0.05. \label{fig:P.808-MOS-results}}
\end{figure}

In addition to the official challenge experiments, we conducted further
experiments on the challenge blind test set. Those experiments were
all conducted after the submission deadline so as to not influence
the model to be submitted. We compared the quality obtained with lower
complexity versions of the proposed algorithm (Section~\ref{sub:sparse-model}).
More specifically, the three RES model sizes were each evaluated with
and without a linear AEC in front. In addition, the AEC alone (no
RES) was evaluated, along with the AEC followed by the SpeexDSP conventional
joint RES and NS. The MOS results from all 600~utterances that include
near-end speech (i.e. excluding far-end single-talk samples) are shown
in Fig.~\ref{fig:P.808-MOS-results}. They demonstrate that the PercepNet-based
RES significantly out-performs the SpeexDSP conventional RES, even
when used as a pure echo suppressor (except for the lowest complexity
setting). Despite the good double-talk performance when operated as
a residual echo suppressor, the results demonstrate the benefits of
using the adaptive filter component. 

\begin{figure}

\centering{\includegraphics[width=1\columnwidth]{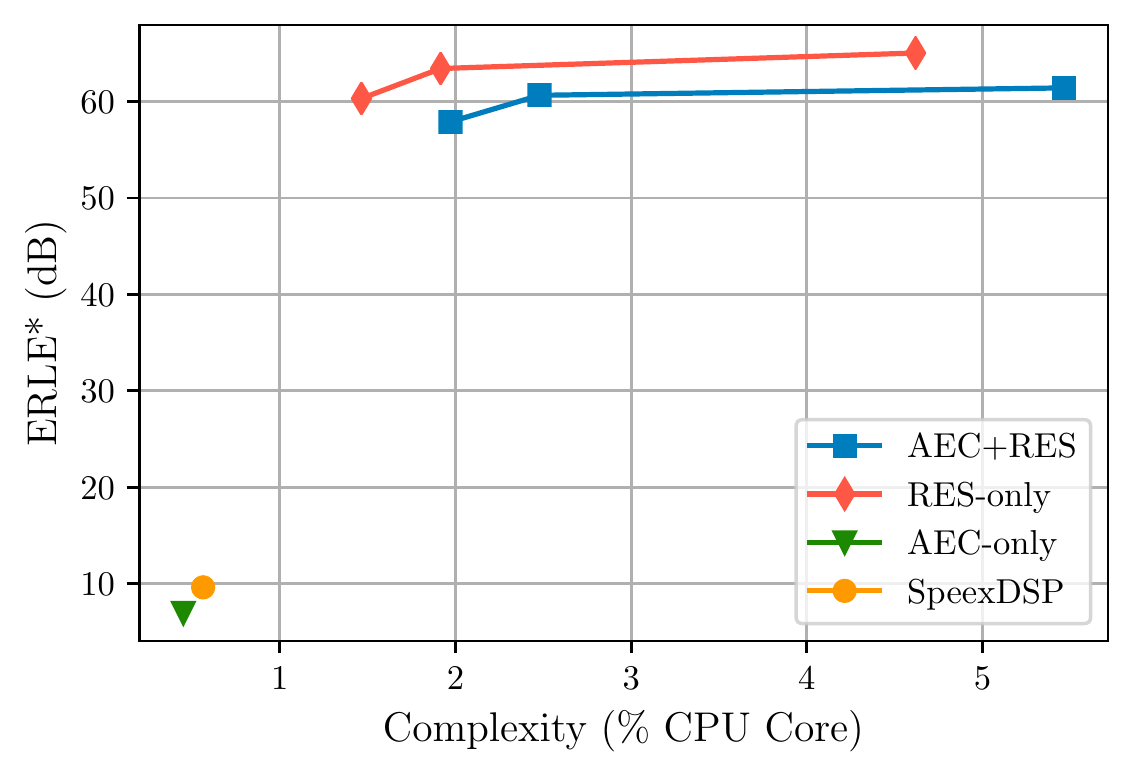}}

\caption{Median ERLE{*} on the far-end single-talk cases as a function of complexity.\label{fig:ERLE-results}}

\end{figure}

The far-end single-talk samples are evaluated based on a modified
echo return loss enhancement (denoted ERLE{*}) metric where both noise
and echo are considered. Since the RES is meant to remove all energy
from those samples, we simply find the ratio of the input energy to
the output energy. The results in Fig.~\ref{fig:ERLE-results} show
that all PercepNet-based algorithms remove far more echo and noise
than the conventional approach. Combined with Fig.~\ref{fig:P.808-MOS-results},
these results confirm that the linear AEC does not help attenuating
isolated (far-end-only) echo, but greatly contributes to preserving
speech during double-talk. 

\section{Conclusion}

\label{sec:conclusion}

We demonstrate an integrated algorithm for echo and noise suppression
in hands-free communication systems. The proposed solution, based
on the PercepNet model, incorporates perceptual aspects of human speech
in a hybrid DSP/deep learning approach. Evaluation results show significant
quality improvements over both traditional and other neural echo control
algorithms while using only 5.5\% of a CPU core. We further evaluate
the impact of the model size on quality down to 1.5\% CPU. We believe
these results demonstrate the benefits of modeling speech using perceptually-relevant
parameters in an echo control task. 

\balance

\bibliographystyle{IEEEbib}
\bibliography{percepnet}

\end{document}